\documentclass[twocolumn,english,eqsecnum,aps,pre,showpacs]{revtex4}
\usepackage[utf8]{inputenc}
\usepackage[english]{babel}

\usepackage{amsmath}
\usepackage{amssymb}

\makeatletter
\usepackage{graphicx}   
\usepackage{dcolumn}    
\usepackage{bm}         
\usepackage{babel}

\begin{document}

\title{Non equilibrium thermal and electrical transport coefficients for hot metals}

\author{J.L.~Domenech-Garret}
\email{domenech.garret@upm.es}
\affiliation{
 Departamento de F\'{\i}sica Aplicada,   
E.T.S.I. Aeron\'{a}utica y del Espacio.\\ 
Univ. Polit\'{e}cnica de Madrid, 28040 Madrid, Spain.}

\date{\today}

\begin{abstract}
This work discusses about the transport coefficients for non equilibrium hot metals. First, we review the role of the non equilibrium  Kappa distribution in which the Kappa parameter varies with the temperature.
A brief discussion compares such distribution with the classical non equilibrium function. Later, we analyze the generalized electrical conductivity obtained from the evolution of the Kappa distribution. Also, we reexamine the connection between a material-dependent coefficient which can be extracted from the thermionic emission and the melting point of  the  metal. We extend previous studies by analyzing additional metals used as thermionic emitters. Finally, in the light of the Wiedemann-Franz Law, we present a new generalized thermal conductivity, which is also applied to several metals.           
 
\end{abstract}

\maketitle

\section{Introduction}
\label{introd}
As it is well known,  metals in equilibrium are described by means of the Fermi-Dirac distribution. However there are many situations in  which the non equilibrium  is present, and it is difficult to assume that this distribution depicts the electron population of such metals. This is the case of metals  rapidly heated by means of dense currents, \cite{Lebedev}. Also, the non equilibrium is present in metals exposed to a high power laser, \cite{Mueller2013}. The deviation from equilibrium  has been also investigated concerning the thermionic emission  \cite{Zakharov, Domenech-G1,Domenech-G2}. Moreover, there are reported departures from equilibrium of metals near their melting point \cite{Lebedev,Domenech-G3}. In addition, the departures from equilibrium of particle distributions have been also applied in wider contexts \cite{Zakharov}.

The Kappa distributions  are used to a large extent in  statistical mechanics and other fields to study populations near and far from equilibrium. The classical Kappa is applied to nonequilibrium  plasmas in  space \cite{Podesta, Shizgal,Livadiotis}. These kinds of distributions can be cast from  general formalisms. The non-extensive q-statistics \cite{Tsallis1,Tsallis2}, and the so-called Beck-Cohen superstatistics 
provide physical meaning to these distributions \cite{Beck}. Besides, Kappa  distributions can be studied from the so called  Kappa-deformed algebras \cite{Kaniadakis}. The non-extensive statistics have been also applied  to establish general purpose Kappa distributions for bosons and fermions \cite{Algin,Treumann}. From another point of view, the Kappa  distribution can be regarded as the solution of the Fokker-Planck equation regarding collective effects and collisional processes,  \cite{Hasegawa}.     

In this work we will apply the Fermi-Dirac Kappa function to study in a phenomenological way the behaviour of the electron population in metals out of equilibrium. We will focus on metals which are used as thermionic emitters. We will consider a small volume of a metal with an external  electric field applied to it. Additionally, such  metal is heated by means of an inner current. This heating can be increased until the melting point. Besides, from the evolution of the Kappa energy distribution has been derived a generalized Ohm law containing a generalized electrical conductivity, \cite{Domenech-G3}. Such conductivity evolves with the temperature, and it drops when such a melting temperature is reached. According to \cite{Lebedev,Domenech-G3}, the starting mechanism of melting can be originated from the enhancement of energy deposition on the metal lattice caused by increasing high energy electron population. Subsequently, it would increase the defect density on the lattice leading to the melting.  

Through this work, we will review this generalized electrical conductivity and we will apply it to other metals. Moreover, as we shall see, from such a generalized coefficient, the Wiedemann-Franz Law can be extended  in a phenomenological manner to the non-equilibrium, giving rise  to a  generalized thermal conductivity. The resulting generalized law is applied later to study the thermal conductivity of several metals.

\begin{figure}
\includegraphics[width=8.5cm]{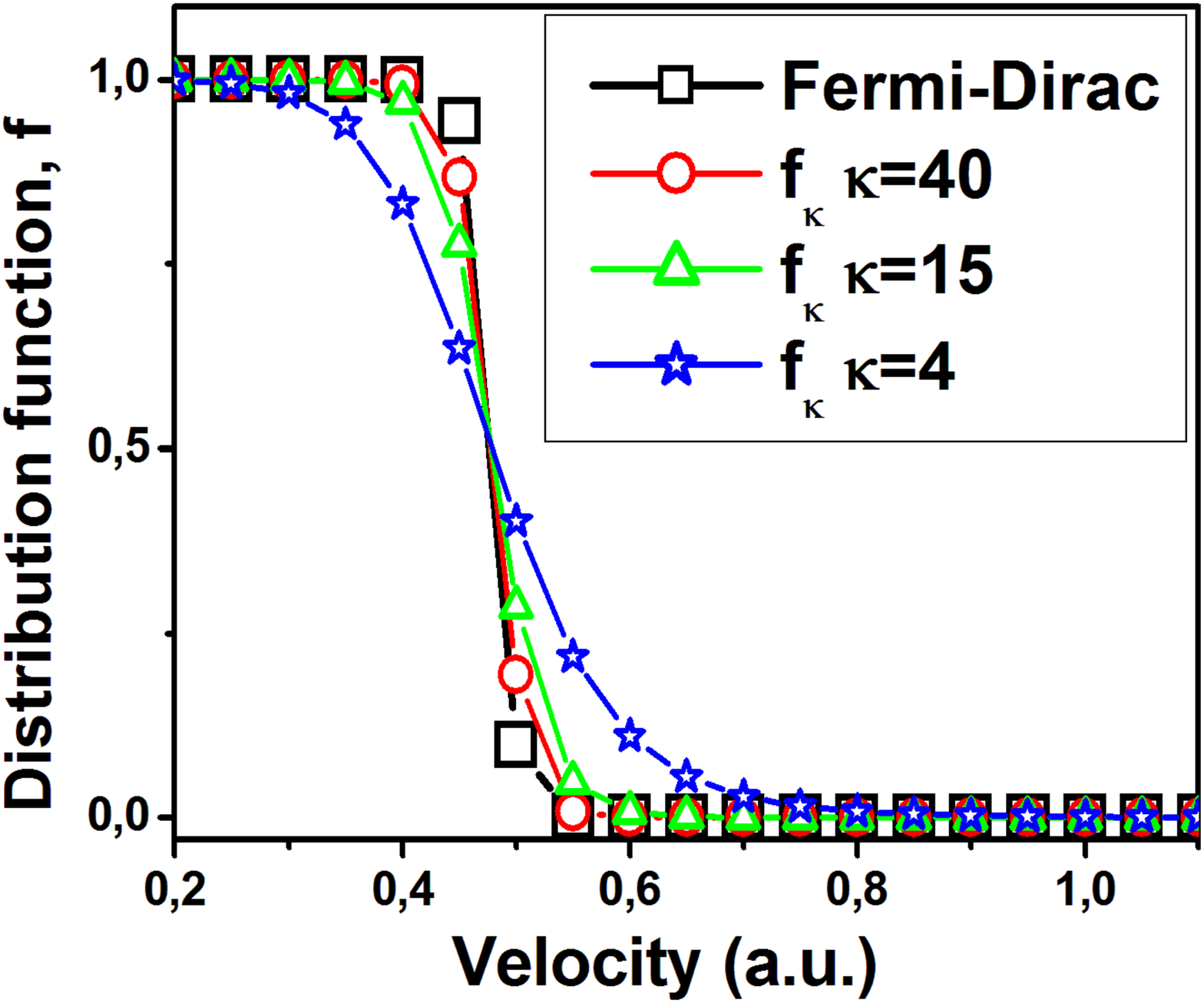}  
\caption{ (color on line) Behaviour of the $\kappa$-FD 
distribution of Eq.(\ref{eq:k-FD}) for different values of $\kappa$. Here, $\phi(\kappa)=(\kappa - 3/2 )\, k_{B} T + \epsilon_{F}$.
The equilibrium Fermi-Dirac distribution is recovered as the Kappa becomes higher.   
\label{fig:k-FD}} 
\end{figure} 
\section{The Kappa distribution in metals}
\label{backg}

The general purpose form for the generalized Kappa Fermi-Dirac distribution can be put as follows \cite{Livadiotis,Treumann},
\begin{equation}
 f_{\kappa}^{ FD} \, ( T, E) 
 =  \left[  1+ \biggl( 1 + \frac{E - \epsilon_{F}}{\phi(\kappa)} \biggr)^{(\kappa+1)} \right]^{-1}
 \label{eq:k-FD}
\end{equation}
\noindent
where $T$ is the temperature, $E$, the electron energy, $\epsilon_{F}$ being  the Fermi energy. Here, $\phi(\kappa)$ is a linear function proportional to $k_{B}T$, with $k_{B}$ being the Boltzmann constant. Such $\phi$ function can be  extracted from a characteristic velocity which is related with the mean square velocity $\left\langle v^2\right\rangle$, \cite{Podesta,Domenech-G1}.  
This distribution, as it can be seen in Figure \ref{fig:k-FD}, is able to develop high energy tails as the Kappa value becomes lower. The FD distribution is recovered for high Kappa values. Moreover, the Fermi energies lie around few $eV$; much higher than the melting temperature of real  metals. Therefore, following reference \cite{Domenech-G1}, from this distribution it can be derived an expression  which is suitable for metals.  The Generalized FD distribution  becomes  smoothed for low  Kappa values when $(E - \epsilon_{F})\ll  k_{B} T$ or $(E - \epsilon_{F})\gg  k_{B} T$. Under such conditions, the following distribution has been built to describe the metals \cite{Domenech-G1}, 
\begin{equation}
\label{fk-e}
f_{\kappa} ( T ,E ) = C_{\kappa}(T) \,  \biggl( 1\ 
+ \frac{E - \epsilon_{F} }{E_{\kappa} + \epsilon_{F}\, } \biggr)^{-(\kappa+1)} 
\end{equation}
\noindent
Here, the  constant $C_{\kappa}(T) $ is normalized to the electron density of the metal, \cite{Domenech-G1}. In this case, the $\phi(\kappa)$ factor of Equation (\ref{eq:k-FD}) takes into account the velocity at the Fermi level $\epsilon_{F}= m v^2_F/2$ within the calculation of the mean square velocity. Then, according to reference \cite{Domenech-G1}, the $\phi(\kappa)$ term   transforms into $E_{\kappa} + \epsilon_{F}$, where $E_{\kappa} = (\kappa - 3/2 )\, k_{B} T$.  As expected,  for high Kappa values corresponding to equilibrium, $f_{\kappa}( T ,E )$  recovers the FD distribution.

\begin{figure}
\includegraphics[width=8.5cm]{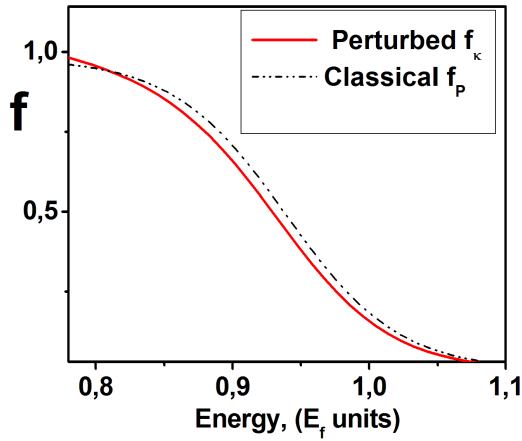}  
\caption{ (color on line) Comparison between the usual perturbed distribution of metals $f_{p}$ (dot-dashed line)  and the linearized $f_{\kappa}$(solid line) : $T=3000K$; $\mathcal{F}=1.9 kV$    
\label{fig:fp-linfk}} 
\end{figure}

In addition, we can  test  the suitability of the Kappa distribution to describe a nonequilibrium metal. We can compare the $f_{\kappa}$ distribution with the classically linearized distribution function usually employed to describe the out of equilibrium electron population of a metal merged within an electric field, $\mathcal{F}$, \cite{Lindsay-X, Ziman1,Domenech-G3}. Under these conditions, the distribution function as altered by the field can be written in terms of the equilibrium Fermi-Dirac distribution. The non-equilibrium distribution reads as,

\begin{equation}
\label{eq:pert-f}
f_p = f_{FD}\ +\ v_x\ g(x,v_x, v_y, v_z) 
\end{equation}
\noindent
in which, the $g$ function describes the perturbed part and it is a function of the velocity components through the modulus of $\mathbf{v}$. 
The latter can be put in terms of the evolution of $f_{FD}$  due to the temperature gradient and the electric field, 

\begin{equation}
\label{eq:gform}
g(x,v_x, v_y, v_z)= -\tau \biggl[ \frac{\partial f_{FD} }{\partial x}\ +\ \frac{e\ \mathcal{F}}{m |\mathbf{v}|}\  \frac{\partial f_{FD} }{\partial |\mathbf{v}|}\biggr]
\end{equation}
\noindent
where $m$ is the electron mass, and $\tau$ stands for a relaxation time, which is related with the electron mean free path, $l=|\mathbf{v}|\, \tau$,  \cite{Lindsay-X,Domenech-G3,Ziman2}. This perturbed term can be related with the  usual BGK collision operator \cite{Domenech-G3, BGK}.
To compare with the first order $f_p$ distribution, we  linearize  the $f_{\kappa}$ distribution, Eq.(\ref{fk-e}), by developing it in the Taylor series.  The result can be seen in Figure \ref{fig:fp-linfk}. Both distributions were evaluated at $T=3000 K$. As the input for $f_p$, the electric field was also inserted. From this Figure we realize the behaviour of both distributions is similar. As it will be discussed later, this Figure also suggest the Kappa parameter should also entangle the electric field.

Moreover, from the above distribution, the thermionic current emitted from the metal surface along the perpendicular direction pointed by the $OX$ axis, can be calculated \cite{Domenech-G1}  using

\[
J_{\kappa} ( T ) = \, e \,  \int^{\infty}_{v_{ox}}  \int^{\infty}_{-\infty}  \int^{\infty}_{-\infty} 
\! \!  v_{x}\ \ f_{\kappa}( T, v )\ d^3\bm{v}    
\]
\noindent
Here  $v_{ox}$ represents the minimum of velocity needed to overtake the potential barrier. This yields an expression which has the form \cite{Domenech-G1},

\begin{equation}
\label{Jk}
J_{\kappa} (T) =\ B_{\kappa} (T)\,  
\times \biggl( 1\ + \frac{W_{f}}{ E_{\kappa}+ \epsilon_{F}} \biggr)^{-\kappa+1}
\end{equation}

\noindent
Here, $B_{\kappa} (T)$  is a factor, and $W_{f}$ is the usual work function. For low Kappa values this expression predicts thermionic currents much higher than those calculated by the Richarson-Dushmann (RD) law. In the limit for large $\kappa$, the above Equation (\ref{Jk}) recovers the RD law. 

To test Equation (\ref{Jk}), in a previous work, \cite{Domenech-G2}, a set of experiments has been done. Briefly, it consists of a Tungsten wire  merged into a vacuum chamber and heated with a DC current, $J$. By means of a sweep system, the emitted current from the wire surface is measured. From these experiments,  a departure from the RD expression can be  observed when  the wire temperature becomes higher. The high values of the measured emitted current at high temperature can not be recovered from the classical expression. The deviation from equilibrium was calculated using Eq.(\ref{Jk}) as a function of the Kappa parameter. As an experimental result, the Kappa parameter was found to be temperature dependent, $\kappa(T)$. The possibility of  a $\kappa(T)$ dependence has been also pointed out in a more general study on suprathermal distributions,  \cite{Treumann}. Using Equation (\ref{Jk}) and the experiments, a linear Kappa law was found through a pure fit of the current along the available range of temperatures,\cite{Domenech-G2}.
\begin{equation}
\label{lin-K-T}
\kappa(T)= a\ -\ b\ T
\end{equation}
\noindent
The $b$ slope was found to be material dependent. For thermionic metals, this value is $b \simeq 10^{-2} (K^{-1})$ \cite{Domenech-G2,Domenech-G3}.  By inserting this $\kappa(T)$ law into the emitted current $J_{\kappa}(T)$, the experimental results are recovered with a very good agreement over the entire temperature range. Therefore, the experiments and this later law, indicate that the departure from the equilibrium of metals is governed by the temperature modulated by the $b$ coefficient. The scope and the suitability of such a linear $\kappa(T)$ law will be discussed later.

\begin{figure}
\includegraphics[width=8.5cm]{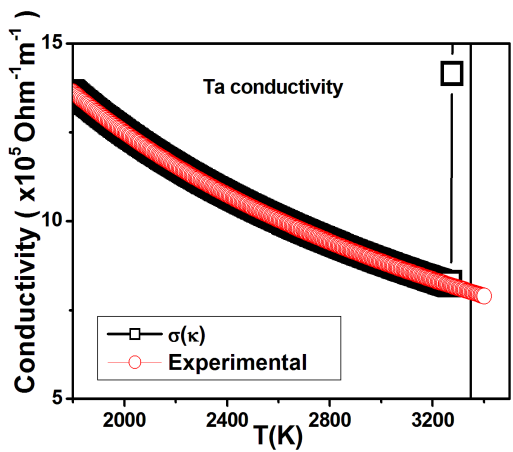}  
\caption{ (color on line)  Electrical conductivity. Comparison of $ \sigma_{\kappa}$ with the experimental data: Case of  Tantalum.    
\label{fig:Ta-sigmak}} 
\end{figure} 

\section{Generalized thermal and electrical conductivity of metals}
\label{Gen}
In this section, we will analyze the impact of the $\kappa(T)$ law when the Kappa distribution evolves according to the Boltzmann Transport Equation (BTE). 
\begin{equation}
\label{Boltz2}
\left[\frac{\partial f }{\partial t}\right]_{coll}=\ v_x\ \biggl[\ \frac{\partial f_{\kappa} }{\partial T}\  \frac{dT}{dx}\ +\ e\ \mathcal{F}\   \frac{\partial f_{\kappa}}{\partial E } \biggr]
\end{equation}
\noindent
In which we use the BTE with $f_{\kappa}(T)$, Eqs.(\ref{fk-e}) and (\ref{lin-K-T}). The calculations lead to a generalized Ohm Law $J=\sigma_{\kappa} \mathcal{F}$ with an electrical conductivity, $\sigma_{\kappa}$,  which can  be written as, \cite{Domenech-G3},

\begin{equation}
\label{ksigma}
\sigma_{\kappa}=\ \sigma\ \biggl[ 1+ \mathcal{C}_{\kappa}(T,E) \biggr]
\end{equation}  
\noindent
Here, $\sigma$ stands for the usual, equilibrium, electrical conductivity. The corrective term is labeled as $\mathcal{C}_{\kappa}(T,E)$, and it becomes zero for large $\kappa$. Therefore the usual Ohm law is recovered for metals in equilibrium.  The explicit form of $\sigma_{\kappa}$ can be read in Equation (2.20) of \cite{Domenech-G3}. As the temperature approaches the melting point, $T_m$, this correction begins to be  important. When $T_m$ is reached,  the $\sigma_{\kappa}$ becomes singular  and subsequently it drops. The responsible for this vanishing conductivity is the singular behaviour of the Digamma function, $\psi[\kappa(b,T)]$, contained within the corrective term. Such Digamma depends on the temperature and the material properties through Kappa. As the temperature rises the Kappa value decreases and it is controlled by the slope $b$. Since such a coefficient is material dependent, this suggests that it contains specific  information about how the particular metal departs from equilibrium, \cite{Domenech-G3}.

If there is available experimental data about the slope of $\kappa(T)$  of a metal from its thermionic emission, it can be used within equation (\ref{ksigma}) in order to predict its melting point. Conversely, using the value of the melting points of metals, $T_m$, as an input,  by finding roots from  equation (\ref{ksigma}) the $b$ coefficient can be obtained, \cite{Domenech-G3}. 

In this work we provide the study  about the $\sigma_{\kappa}$ and $b$  for Tantalum, since it is also used as a thermionic emitter, \cite{Cardarelli}.  Figure \ref{fig:Ta-sigmak} shows the comparison of the experimental behaviour of the electrical conductivity of Tantalum  with the temperature and the obtained values from  equation (\ref{ksigma}). The experimental data relevant to this study was collected  from \cite{Cardarelli,CRC,Desay}. The  calculated coefficient  is $ b=0.02\ K^{-1}$  from its melting point $T_m$(Ta)$= 3293 K$. The typical Kappa values  far from the melting point are $\kappa( T=1500 K ) = 39$; a high value which would correspond to a distribution almost in equilibrium. Near the  melting $\kappa( T=3180 K ) = 5$, it shows the transition toward non-equilibrium. These results are similar to those found in other thermionic metals in reference \cite{Domenech-G3}. The electron population develops high energy tails interacting with the metal lattice and increases the disequilibrium until it reaches the melting point. This interaction leads to the enhancement of  energy deposition on the particular metal lattice. Subsequently this absorption increases the defect density in the lattice \cite{Lebedev}, leading to melting.

\begin{figure}
\includegraphics[width=8.5cm]{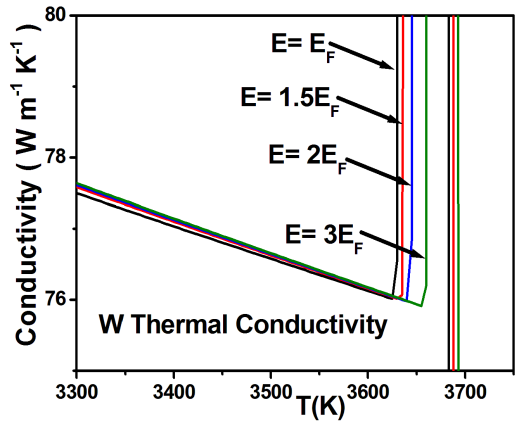}
\caption{ (color on line)  Thermal conductivity, $ \chi_{\kappa}$ of Tungsten for different energies in Fermi energy  units, $E_F$.    
\label{fig:W-Chik}} 
\end{figure} 

On the other hand, we can use the Equation (\ref{ksigma}) to generalize the Wiedemann-Franz law (WFL). First, we define the generalized thermal conductivity, $\chi_{\kappa}$, 

\begin{equation}
\label{kchi}
\chi_{\kappa}=\ \chi \biggl[ 1+ \mathcal{C}_{\kappa}(T,E) \biggr] 
\end{equation}  
\noindent             
where $\chi$ stands for the usual thermal conductivity of metal in equilibrium. The $\mathcal{C}_{\kappa}$ within brackets corresponds to the same corrective term of Equation (\ref{ksigma}).\newline
The classical WFL reads as,

\begin{equation}
\label{WFL}
\frac{\chi}{\sigma\ \ T}=L_N\ 
\end{equation}  
\noindent
where,  again, $\sigma$ is the equilibrium electrical conductivity, and $L_{N}$ is the Lorentz number.  We can rewrite the WFL in terms of the generalized electrical conductivity. Hence, by inserting   Equations (\ref{ksigma}) and (\ref{kchi}) into the above expression we attain,

\begin{equation}
\label{kWFL}
\frac{\chi_{\kappa}}{\sigma_{\kappa}\ \ T}=L_N
\end{equation}  
\noindent
Therefore, the same  WFL holds by replacing the usual transport coefficients with their respective generalized expressions. Also, from  Equation (\ref{kWFL}) it is possible to calculate $\chi_{\kappa}$ through a knowledge of the generalized electrical conductivity of a particular metal.

\begin{figure}
\includegraphics[width=8.5cm]{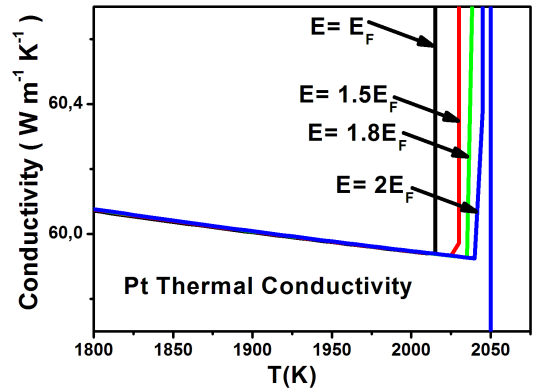}  
\caption{ (color on line)  Thermal conductivity , $ \chi_{\kappa}$ of Platinum for different energies in Fermi energy  units, $E_F$.     
\label{fig:Pt-Chik}} 
\end{figure}

In Figures \ref{fig:W-Chik}  and \ref{fig:Pt-Chik},  we can observe the results of $\chi_{\kappa}$ obtained for Tungsten and Platinum using Eqs.(\ref{ksigma})
and (\ref{kWFL}) for different energies, in $\epsilon_F$ units. In the first case the $\sigma_{\kappa}$ were obtained from the experimental $b$ coefficient for Tungsten. In the Pt case the corresponding $\sigma_{\kappa}$ were calculated from the experimental melting point. Both cases agree with the available experimental  extrapolations. These extrapolations are based on different experiments on thermal conductivity near the melting point, \cite{Powell}. As a new feature, the predicted thermal conductivity of the solid phase of the metal drops when the melting point is reached.     
    
\section{Conclusions}

In this work we have studied the previously  modeled \textit{ad-hoc} Kappa distribution for metals, which comes from the general Kappa Fermi-Dirac. This $f_{\kappa}$ function is normalized to the electron density of the metal. By analyzing the thermionic emission of several metals, it was found that the Kappa parameter depends on temperature. From the experiments, this $\kappa(T)$ law was found to be linear. At this point it must be noticed that this linear law was extracted in a phenomenological way within a limited temperature range, not from theory. Therefore, the possibility to find this law as an effective expression constrained to a small temperature range can not be discarded.

The $b$ slope actually measures the rate of change of $\kappa$ with respect to the temperature and it could entangle the physics of such a mechanism.  In addition, to analyze the  suitability of $f_{\kappa}$ for metals, we compared it with the classical perturbed distribution function for metals, finding a similar behaviour. This latter function depends on the electric field applied to the metal. As it is well known such a field is a source of disequilibrium \cite{Lindsay-X,Ziman1}. Therefore it supports the idea the $\kappa(T)$ law is an effective expression. It is expected the Kappa parameter depends not only on the temperature, but it should be also a function of the applied electric field. Looking at the linear $\kappa(T)$ law, it suggests the $b$ coefficient entangles information not only about the features of the metal, but also the conditions leading to nonequilibrium. Hence, it is mandatory to obtain an explicit expression of this  $\kappa(T,\mathcal{F},...)$ law. This work is in progress.

Concerning the mechanism of the development of  high energy tails, again it is entangled within the \textit{b}  coefficient of the $\kappa(T)$ index. The phase transition is mainly described by the Digamma terms within $\mathcal{C}_{\kappa}(T,E)$. As mentioned, the Kappa distributions develop through the adjustable Kappa parameter. Such an index is modeled on the terms of the  wave-electron interactions and strengths, \cite{Shizgal}. Near  the phase transition it is expected that the lattice oscillations will become larger. Their interactions with the electron-lattice waves could  be the origin of the development of such (low-Kappa) high energy tails. Similar effects analyzing the interaction between the phonons  and electrons in heterolayers have been reported \cite{Kiwook}. In this latter case the cause of the increasing energy of the electron distribution comes from the reabsortion of non-equilibrium phonons. In our case, as the temperature rises, the increasing defect density in the metal lattice  could play the role of the heterointerfaces. Again, more effort is needed to clarify such a topic.

On the other hand, in this work  we extended previous studies by applying the generalized expression of the electrical conductivity to the case of Tantalum. This latter metal is also used as a thermionic emitter. We found all the values extracted are consistent with previous results from other metal emitters.  Finally, in  light of the Wiedemann-Franz Law, we presented a new generalized thermal conductivity. This Kappa transport coefficient led us to generalize the WFL to non-equilibrium. As well, this generalized WFL allowed us to calculate the thermal conductivity values near the melting point of several metals. These values are consistent with the experimental extrapolations of the thermal conductivity  near their melting points. In addition, as a new feature, such an expression makes the thermal conductivity  vanish when the melting point of the solid phase is reached.

\begin{acknowledgments}
This work was funded by the MINECO, Spanish Ministry, under Grant ESP2013-41078-R.
\end{acknowledgments}


\end{document}